\documentclass[epsf,twocolumn,showpacs,floatfix,preprintnumbers]{revtex4}
\usepackage{graphics}
\usepackage{graphicx}
\usepackage{dcolumn} 
\usepackage{bm}
\usepackage{color}
\usepackage{epsfig}
\begin{document}
\title{Compensated Half-metallicity in the Trigonally
Distorted Perovskite-type NiCrO$_3$
}
\author{Kwan-Woo Lee$^1$ and Warren E. Pickett$^2$} 
\affiliation{ 
$^1$  Department of Display and Semiconductor Physics, 
  Korea University, Jochiwon, Chungnam 339-700, Korea\\
$^2$Department of Physics, University of California, Davis,
  California 95616, USA
}

\date{\today}
\begin{abstract}
Using first principles calculations, we investigate 
the electronic and magnetic properties of
the trigonally distorted ($R\bar{3}c$) perovskite-derived NiCrO$_3$.
Within the local spin density approximation (LSDA), our calculations
show that this system is an exactly compensated half-metal (CHM).
The local spin moments of Cr 2.04, and antialigned Ni --1.41 and 
three oxygens --0.63 (in the units of $\mu_B$),
indicate high spin $S=\frac{3}{2}$ Cr$^{3+}$ and 
$S=\frac{3}{2}$ (NiO$_3$)$^{3-}$ units.
Considering reasonable values of the on-site Coulomb repulsion $U$ 
on both Ni and Cr ions with LDA+U approach, 
this system becomes an insulator (as reported by Chamberland and
Cloud) having a narrow gap
in the spin-up channel, whereas the other channel
has a large gap of $\sim$3 eV.
Although inclusion of $U$ seemingly leads to the transition
Ni$^{2+}$$\rightarrow$high spin $S=\frac{3}{2}$ Ni$^{3+}$,
consistent with the experimentally observed effective moment,
the zero net moment remains unchanged due to either reduction of
oxygen local moments or enhancement of Cr local moment. 
Compression of volume by 10\% leads to
CHM even when correlation effects are included.
These results suggest the possibility of a CHM state in NiCrO$_3$ and
provide another route to search for CHM, which is a property sought by many.
\end{abstract}
\maketitle

\begin{figure}[tbp]
{\resizebox{3.5cm}{7cm}{\includegraphics{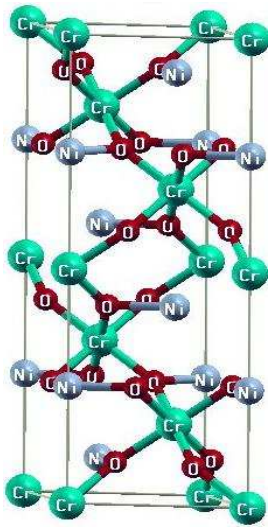}}}
\caption{(Color online) Crystal structure of the trigonally distorted
NiCrO$_3$, having the corner-sharing CrO$_6$ octahedra.
}
\label{str}
\end{figure}

{\it Introduction.}
A half-metallic antiferromagnet, more properly called a compensated half-metal (CHM), 
is half-metallic, having one conducting spin channel and one insulating
spin channel, but vanishing (compensating) net magnetic moment, hence no 
macroscopic magnetic field.
Since proposed by van Leuken and de Groot\cite{leuken} fifteen years ago,
the CHM has been considered as an encouraging candidate for spintronics -- 
fully polarized carriers without a macroscopic magnetic field to deal with --
and a possible single spin superconductor.\cite{wep1}
However, a true CHM has not yet been established.

Since at least two different kinds of magnetic ions 
are necessary for vanishing moment,
the double perovskite systems were investigated
in several theoretical investigations.\cite{wep2,uehara,park,LP1,PP,song,chen}
However, preparing an ordered sample has been frequently failed.\cite{androu,jana} 
The Heusler structures, which have produced many half-metals,
also have been considered as another promising 
candidate.\cite{wep3,felser,erosy}
Nakao has suggested CHM in tetrahedrally coordinated chalcopyrites\cite{nakao1}
and in their monolayer superlattices.\cite{nakao2}

In this study, we will extend the search for CHMs to the trigonally 
distorted perovskite-related systems having two different transition metals.
Several systems of this class were synthesized about forty years
ago.\cite{chamb,chamb2,sawamoto}
Here we will investigate NiCrO$_3$ with a
corundum-related structure, 
synthesized by Chamberland and Cloud.\cite{chamb}  Their procedure involved 
preparation at high (60-65 kbar) pressure at 1200 $^\circ$C, 
followed by quenching to
room temperature, so intermixing on the Ni and Cr sites cannot be ruled out.
The susceptibility measurements showed a Curie-Weiss behavior in the high
$T$ regime and discontinuities at 250 K and 120 K, 
proposed to be canted antiferromagnetism.
The value of the effective moment was $p_{eff}$=5.8 $\mu_B$, close to 
the value for Cr$^{3+}$ and Ni$^{3+}$,  both carrying high spin $S=\frac{3}{2}$,
implying a possibility of CHM.
This system was proposed to be semiconducting with an activation energy
of 0.11 eV, though no detailed information was provided.
More recently, this compound has been obtained as a minority
during the oxidization process by a few groups,\cite{leyens,gorhe} 
who measured only the space group and lattice constants consistent with
those of Chamberland and Cloud.\cite{chamb}

{\it Structure and methods.}
In this trigonally distorted system, Chamberland and Cloud reported only 
the space group $R\bar{3}c$ (No. 167) with the hexagonal lattice constants 
$a=4.9252$ \AA~ and $c=13.504$ \AA,\cite{chamb}
leading to a distortion factor $\frac{a\sqrt{6}}{c}$ of 0.89
(unity for the ideal perovskite).
We have optimized the structure parameters in the cubic and the trigonally 
distorted structures.
In both structures, the one in which the Ni ion sits on the A site and the Cr ion  
on the B site is much favored over reversed site occupation. 
Consistent with the experiments, our optimizations show
that the cubic phase with optimized lattice constant $a=3.602$ \AA~ 
has much higher energy, by a few eV, than the distorted phase.
Hence we will consider only the distorted structure.

In the $R\bar{3}c$ structure, 
using the experimental lattice parameters,\cite{chamb}
our optimizations indicate that Ni, Cr, and O ions lie 
on the $6a$ (0,0,$\frac{1}{4}$), $6b$ (0,0,0), 
and $18e$ ($x$,0,$\frac{1}{4}$) sites, respectively,
with the oxygen internal parameter $x$ of 0.3748.
As displayed in Fig. \ref{str}, 
the Cr-O bond length of 1.915 \AA~ is about 3\%
larger than the Ni-O bond length of 1.85 \AA.
The CrO$_6$ octahedra consist of a nearly square base and equilateral
triangular faces, with only 1.8\% difference between the O-O bond lengths. 
Thus the CrO$_6$ unit remains close to an ideal octahedron. 
The O-O-O bond angles are 60$^\circ$, 2$\times$60.6$^\circ$, and
58.8$^\circ$. The O-Cr-O bond angles are either 91$^\circ$ or 89$^\circ$.
The Ni ions lie at the center of mass of the equilateral triangle
framed by in-plane oxygens.
This trigonal planar symmetry $D_{3h}$ of the NiO$_3$ leads to
the crystal field splittings: doublet $e_{1g}$ ($d_{yz},d_{zx}$),
doublet $e_{2g}$ ($d_{xy},d_{x^2-y^2}$), and singlet $a_{1g}$ ($d_{z^2}$).

These calculations were carried out within the local spin density approximation (LSDA) 
followed by the LDA+U approach, as implemented in the all-electron full-potential 
code FPLO-9.\cite{fplo}
In the LDA+U approach, two popular double-counting schemes,
the so-called around mean field (AMF) and the fully localized limit 
(FLL),\cite{ldau1,ldau2} were used, showing similar results for this compound.
In our calculations, a rhombohedral cell containing two formula units
was used.
The convergence was checked with a regular mesh containing up to
781 irreducible $k$ points.

\begin{figure}[tbp]
{\resizebox{8cm}{6cm}{\includegraphics{Fig2a.eps}}}
\vskip 8mm
{\resizebox{5.3cm}{4.5cm}{\includegraphics{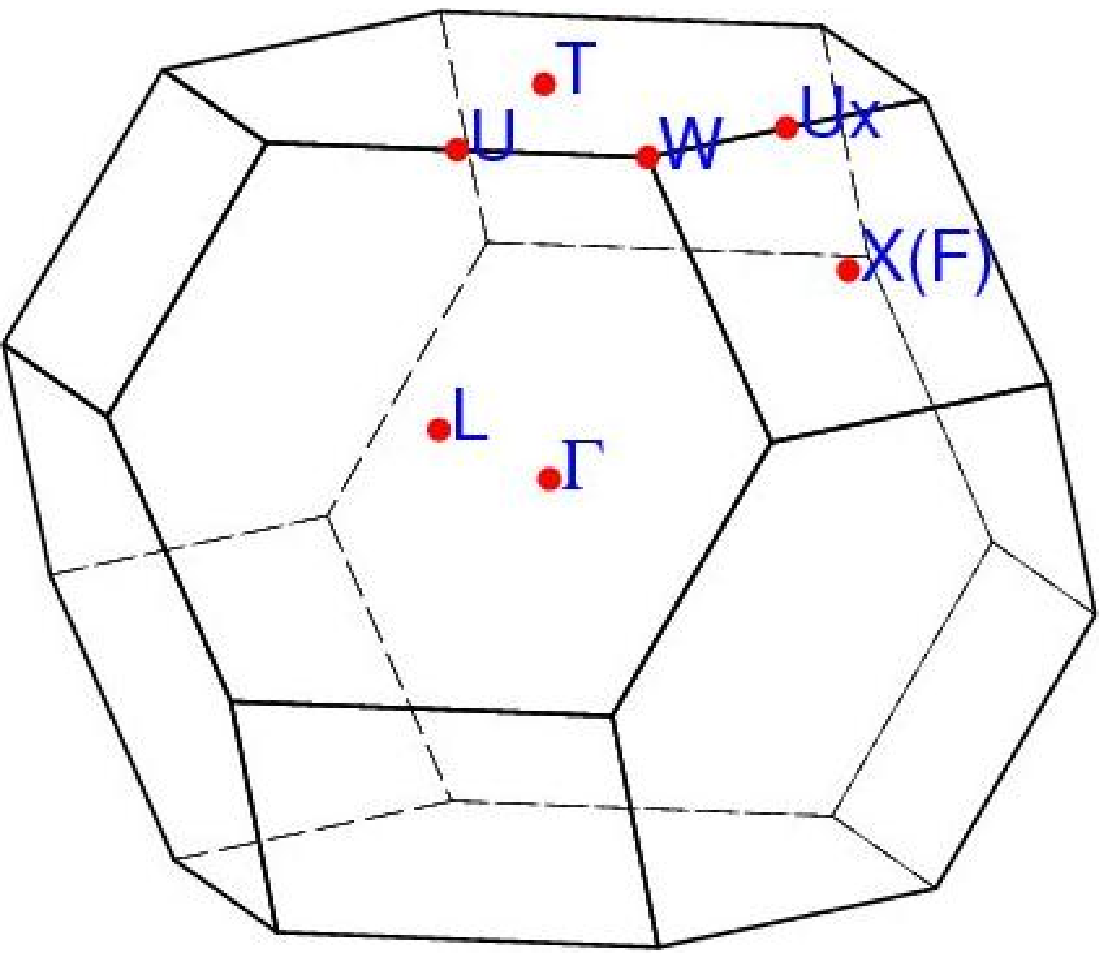}}}
\caption{(Color online) Top: LSDA band structure near the
 Fermi energy $E_F$, where the bands arise only from
 $d$ orbitals of Ni and Cr ions.
 The dashed horizontal line indicates $E_F$, which is 
 taken as the zero of energy.
Bottom: Brillouin zone and high symmetry points of the
 rhombohedral structure, which has similarity to the fcc
 Brillouin zone with a threeefold axis oriented along the $z$ axis.
}
\label{ldaband}
\end{figure}

\begin{figure}[tbp]
{\resizebox{8cm}{6cm}{\includegraphics{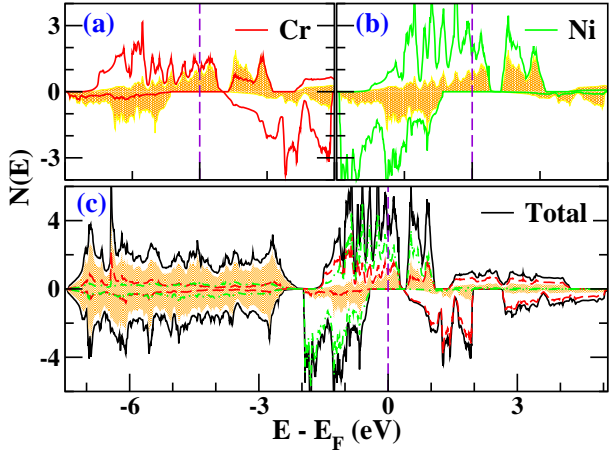}}}
\caption{(Color online) Total and atom-projected LSDA densities of states,
which shows the half-metallic aspect, expressed per eV-formula unit.
In all panels, the shaded regions indicate three oxygens-projected DOS.
The total density of states $N(E_F)$ at $E_F$ is 3.80 states per eV
per a formula unit, decomposed of 50\% Ni, 30\% Cr, and 20\% three
oxygens characters.
(a) Cr $3d$ DOS in solid line, from -2 eV to 2 eV; (b) Ni $3d$, on same
scale as in (a); (c) total DOS in the entire O $2p$ - metal $3d$
region.
The unoccupied Cr $e_{g}$ states with the exchange splitting of 1.5 eV
lie on 1.5--4 eV and 2.5--5 eV in the spin-up and spin-down channels,
respectively.
The exchange splitting of Cr $t_{2g}$ states is about 2 eV.
The $t_{2g}$-$e_{g}$ crystal field splitting of Cr in the spin-up
channel is 3.5 eV, 1 eV larger than in the spin-down channel.
The vertical dashed line indicates $E_F$.
}
\label{ldados}
\end{figure}

{\it LSDA electronic structure.}
The Cr local moment (2.04 $\mu_B$) is precisely canceled by the moments of
antialigned Ni (--1.41 $\mu_B$) and three oxygens (--0.63 $\mu_B$).
The oxygen moment is unusually large
for a $3d$ system, though not unprecedented.
(In the PBE generalized gradient approximation,\cite{gga} 
the CHM configuration also results but with somewhat larger 
moments: Cr 2.26 $\mu_B$,  
Ni --1.54 $\mu_B$, three oxygens totaling to --0.72 $\mu_B$ .)
This CHM state with zero net moment is favored by 300 meV/f.u. over 
the nonmagnetic state.
Thus NiCrO$_3$ is a compensated half-metal within LSDA,\cite{BA}
regardless of the choice of exchange-correlation functional.

The LSDA band structure near $E_F$ is displayed in Fig. \ref{ldaband},
illustrating the 0.8 eV gap in the spin-down channel
leading to the half-metallic state.
In the spin-up (metallic) channel, the Ni $e_{2g}$ bands 
lying in the range of 0.6--1 eV are separated from the mixture of
the other Ni $d$ and Cr $t_{2g}$ bands spread out over the regime
of --1.8 to 0.4 eV. 
In the spin-down bands, the Ni $3d$ (majority) states are fully occupied
with small (1.4 eV) bandwidth, and 
Cr $3d$ (minority) states are empty.
The corresponding densities of states in the full energy range 
and with enlargements emphasizing Ni and Cr separately, are displayed in Fig. \ref{ldados}.
The oxygen bands in the range of --7.5 to --2 eV
are separated from the $d$ bands by 0.5 eV.  However, the mixture of O $2p$
character into both Ni and Cr bands indicates
strong $p-d$ hybridization.

The spin-up channel is more interesting, being a mixture of Cr majority and
Ni minority states.  To balance the five occupied Ni (majority) spin-down bands
(per f.u.) there must be five spin-up bands occupied (on average through the zone, 
due to the metallic nature). Three bands, around 0.5 eV and above, are separated
from the other intermixed seven bands, so the filling is 5/7. These can be
identified as Cr $e_g$ and a substantial admixture of Ni (and of O $2p$ as noted
earlier). The occupied spin-up
bands are Cr $t_{2g}$ admixed with Ni $d_{yz},d_{zx}$, 
and $d_{z^2}$ bands.  Being strongly hybridized and metallic, formal valences
become a murky concept. $S=\frac{3}{2}$ Cr$^{3+}$ seems to be the clearest
feature; presuming O$^{2-}$ ions with their formal valence one would obtain
$S=\frac{3}{2}$ Ni$^{3+}$.
We emphasize that the strong $p-d$ hybridization and metallicity make these
assignments of limited use.

\begin{figure}[tbp]
{\resizebox{8cm}{6cm}{\includegraphics{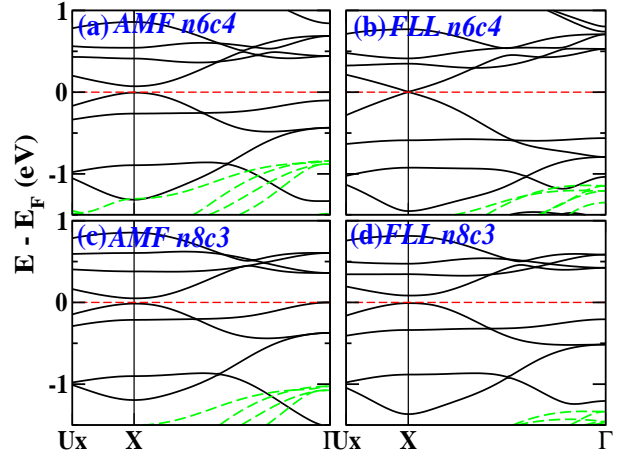}}}
\caption{(Color online) Blowup band structures along the $Ux$-$X$-$\Gamma$
line near $E_F$ for various values of $U$ in both schemes of LDA+U.
Here, $n6c4$ denotes $U$=6 and 4 eV for Ni and Cr ions, respectively.
In panel (b), the gap of 5 meV separates nearly degenerate linear bands 
similar to what has been observed (at the $\Gamma$ point) in CoSb$_3$.[\onlinecite{cosb3}]
}
\label{uband}
\end{figure}

\begin{figure}[tbp]
\vskip 8mm
{\resizebox{8cm}{6cm}{\includegraphics{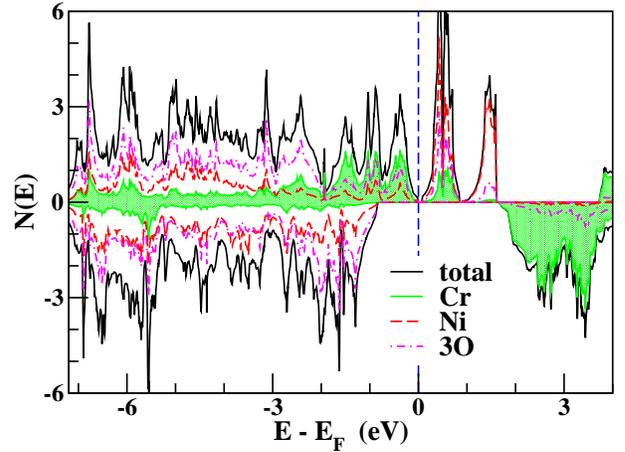}}}
\caption{(Color online) Total and atom-projected densities of states
per formula unit in the AMF scheme of LSDA+U 
using $U$= 4 and 6 eV for Cr and Ni ions, respectively.
Near $E_F$ the occupied part has mostly Cr $d$ character, but
the bottom of the unoccupied part has Ni $d$ character
even though the strong O characters also exist.
In particular, the O characters are nearly the same as
the $d$ character in the range of $-0.1$-- 0.1 eV.
This indicates a nearly compensated half-semimetal.
}
\label{udos}
\end{figure}

{\it Inclusion of correlation effects.}
We have used the LDA+U approach to model the observed insulating state.
The proper $U$ values for this system are unclear, but
values that have been widely used in oxides are 3--4 eV for Cr\cite{LP2} 
and 6--8 eV for Ni,\cite{beng}
having been obtained from comparison with experimental data or calculated from the 
constrained LDA approach.\cite{solovyev}
The Hund's exchange integral $J$=1 eV has been fixed for all values of $U$,
after some variation of $J$ indicated results are insensitive to values 
around this one.

Figure \ref{uband} displays the enlarged band structures
along the $Ux$-$X$-$\Gamma$ line (which is where the gap opens) 
near $E_F$ (set to zero), for these ranges of $U$ and using 
both commonly used LDA+U schemes.  The energy gap
in the spin-down channel increases to
about 2.6 eV in AMF and 3.6 eV in FLL (not shown in the figure).
In the spin-up channel, the bottom of the conduction bands and
the top of the valence bands both consist of the mixture of the 
O $p_z$ and Ni $e_{2g}$ bands, in particular near the $X$ point,
with little Cr $d$ character.
However, about 50\% at the bottom of the conduction bands and the top
of the valence bands are oxygen and Ni characters, respectively.
The second highest valence band, which is nearly flat,
is one of the Cr $t_{2g}$ bands.
In AMF, at $n6c4$ (denoting $U_{Ni}=6$ eV, $U_{Cr}=4$ eV) 
a small gap in the spin-up channel
opens at the $X$ point, whereas
in FLL two bands almost touch at $E_F$.
For $n8c3$, in FLL the band structure is similar with that of $n6c4$,
except for a slight increase in the energy gap.
In AMF, however, Cr $t_{2g}$ bands at the $\Gamma$ point are raised,
leading to an small indirect gap of the $\Gamma$ to $X$ point almost
equal to the direct gap at $X$.
This indirect gap is observed only in the case of $U$=3 eV for Cr ions
in AMF.

The FLL and AMF methods are known to encourage magnetism differently.\cite{erik}
Here we find that applying $U$ on the metal atoms leads to significantly
different induced moments on the oxygen ions.
In FLL, applying $U$ enhances the magnitudes of all local moments, 
and the combined moment of three oxygens is still about 40\% of the Ni moment.
In AMF, on the other hand, both Ni and Cr moments are increased by
applying $U$, but the moment of three oxygens becomes only 20\% of the 
Ni moment.
These different moments can be ascribed to an 
energy penalty for magnetism in AMF that is not present in FLL,
as discussed previously.\cite{erik}
On balance, however, the moment compensation remains unchanged 
when varying $U$ in the range studied here. 

Since differences in the overall DOS curves are small, we show  
only that for $n6c4$ in AMF in Fig. \ref{udos}.
In the spin-up channel, the Fermi energy lies between 
the occupied Cr $d$ bands and the unoccupied Ni $e_{2g}$
bands. Just above 1.5 eV, narrow unoccupied Ni $a_{1g}$ bands,
{\it i.e.} mixing slightly with O states, exist.
As a result, in the insulating phase, the Ni ion is apparently trivalent,
even though the local moment of --1.90 $\mu_B$ is considerably 
reduced due to strong $p-d$ hybridization.
However, the charge difference of Ni ions, obtained
by the Mullikan charge decomposition, between in LSDA and
LDA+U is only $\sim$0.2e, similar to that observed in charge disproportionated
Na$_{1/2}$CoO$_3$.\cite{LP3}

{\it Discussion and Summary.}
Given the very small gap in the up channel, modest pressure
should close the gap, giving a CHM state, as we now show.
At 10\% compression in volume,
the zero moment is slightly degraded within LSDA, but
inclusion of $U$ in the same range as in the ambient pressure
restores the state of the zero moment.
In the spin-up channel, which is insulating at ambient pressure, 
Ni $a_{1g}$ bands of $W \approx$0.6 eV cross $E_F$ 
along the $L$-$\Gamma$-$X$ line, resulting in being
metallic and a precise compensated half-metal. 
(Below $\sim$ 10\% compression, the electronic structures
are very similar with those at ambient pressure.)
This suggests that applying pressure will lead to a CHM state even though
this system is insulating at ambient pressure.

In summary, we have investigated $R\bar{3}c$-structure NiCrO$_3$,
using both LSDA and LDA+U approaches.
Within LSDA, this system is a precise compensated half-metal
with the energy gap of 0.8 eV in the spin-down channel.
Considering the on-site Coulomb repulsion $U$ of both Ni and Cr ions,
we obtained an insulating phase, but retaining zero net moment regardless of
the magnitude of $U$.
Interestingly, the energy gap is substantial (2.6 -- 3.6 eV) in the spin-down channel,
whereas it is tiny or vanishing in the spin-up channel.
Application of pressure provides a likely route to a CHM state in this compound.
These findings provide another structure class that may be favorable for
synthesis of a compensated half metal.

K.W.L. was supported by the Basic Science Research Program through
the National Research Foundation of Korea funded by the Ministry
of Education, Science, and Technology under Grant No. 2010-0008779.
W.E.P was supported by DOE under Grant No. DE-FG02-04ER46111.

\end{document}